\documentclass{article}

\usepackage{PRIMEarxiv}
\usepackage{multirow,tabularx}
\usepackage[utf8]{inputenc} %
\usepackage[T1]{fontenc}    %
\usepackage{hyperref}       %
\usepackage{url}            %
\usepackage{booktabs}       %
\usepackage{amsfonts}       %
\usepackage{nicefrac}       %
\usepackage{microtype}      %
\usepackage{lipsum}
\usepackage{fancyhdr}       %
\usepackage{graphicx}       %
\graphicspath{{figures/}}     %
\usepackage{subcaption}
\usepackage{placeins}
\usepackage{soul,color}
\title{Discovery of 2D materials using Transformer Network based Generative Design
\thanks{\textit{\underline{Citation}}: 
\textbf{Authors. Title. Pages.... DOI:000000/11111.}} 
}

\author{
  Rongzhi Dong \\
 Department of Computer Science and Engineering\\
  University of South Carolina\\
  Columbia, SC 29201 \\
   \And
  Yuqi Song\\
 Department of Computer Science and Engineering\\
  University of South Carolina\\
  Columbia, SC 29201 \\  
  \And
  Edirisuriya M. D. Siriwardane\\
 Department of Physics\\
  University of Colombo\\
Colombo 00300, Sri Lanka \\  
   \And
 Jianjun Hu *\\
 Department of Computer Science and Engineering\\
  University of South Carolina\\
  Columbia, SC 29201 \\
  \texttt{jianjunh@cse.sc.edu} \\
}

\begin{document}
\maketitle

\begin{abstract}

Two-dimensional (2D) materials have wide applications in superconductors, quantum, and topological materials. However, their rational design is not well established, and currently less than 6,000 experimentally synthesized 2D materials have been reported. Recently, deep learning, data-mining, and density functional theory (DFT)-based high-throughput calculations are widely performed to discover potential new materials for diverse applications. Here we propose a generative material design pipeline, namely material transformer generator(MTG), for large-scale discovery of hypothetical 2D materials. We train two 2D materials composition generators using self-learning neural language models based on Transformers with and without transfer learning. The models are then used to generate a large number of candidate 2D compositions, which are fed to known 2D materials templates for crystal structure prediction. Next, we performed DFT computations to study their thermodynamic stability based on energy-above-hull and formation energy. We report four new DFT-verified stable 2D materials with zero e-above-hull energies, including NiCl$_4$, IrSBr, CuBr$_3$, and CoBrCl. Our work thus demonstrates the potential of our MTG generative materials design pipeline in the discovery of novel 2D materials and other functional materials.

\end{abstract}


\section{Introduction}

Two-dimensional (2D) materials have been emerging as a promsing functional materials with wide applications due to the novel fundamental physics with the reduced dimension\cite{glavin2020emerging} The systematic discovery and synthesis of functional 2D materials has been the focus of many studies\cite{2DMatPedia2,lyngby2022data,V2DB,C2DB2,MC2D,wyss2022large,ares2022recent}. 
Having exceptional and tunable properties, 2D materials hold strong promise in semiconductor, energy, and health applications\cite{briggs2019roadmap,li2020new}. Since the 2010 Nobel prize-winning discovery of graphenes\cite{graphene} with a simple 2D structure of carbon atoms but with attractive and complex physics, only a few thousands of distinct 2D materials have been successfully synthesized\cite{C2DB2}. The isolation of single graphene sheets, which proves that 2D systems can exist, gives rise to the discovery of many 2D materials with unique superconducting\cite{cao1}, electronic\cite{manzeli20172d}, magnetic\cite{huang2021two}, and topological properties\cite{kou2017two}. In addition to being test beds for studying the behavior of systems in reduced dimensions, 2D materials hold great promise for various applications in optoelectronics\cite{wang2012electronics}, catalysis\cite{deng2016catalysis}, and the energy sector\cite{anasori20172d}. The research effort has been mainly concentrated on the systems which have bulk counterparts representing anisotropic crystals with layers held together by van der Waals (vdW) forces, with the most prominent example being the graphene and graphite. The weak interlayer interaction leads to a natural structural separation of the 2D subunits in the crystals, therefore making the mechanical or liquid-phase exfoliation possible.

Currently, there are three approaches for generating 2D materials: the top-down exfoliation method starts with a bulk material and exfoliates to make it thinner and peel the layers to obtain 2D materials; the bottom-up approach instead starts with existing 2D materials and uses element substitution to generate new materials. The third one is the de novo structure generation approach\cite{lyngby2022data} based on deep learning generative models such as CDVAE\cite{CDVAE}. 
To get new 2D materials through the exfoliation method, we need to judge whether the 3D bulk material is layered so that it can be exfoliated. The layer screening process first checks the distance between atoms to identify whether these atom pairs are bonded. It then calculates the bonded atom clusters both in a 3x3x3 supercell and the unit cell. If the number of clusters in the supercell is three times that in the unit cell, the structure is tagged as layered\cite{larsen2019definition}. 2D materials are theoretically exfoliated by extracting one cluster in the standard conventional unit cell of the screened layered bulk structures. In the element substitution method, all the elements of the periodic table are categorized into different groups according to their column number. Elements with the same column (group) number share the same number of electrons in their outermost orbit, and elements with the same row (period) number share the same number of electronic layers, which means that elements in the same group or neighbor share some similar chemical properties. The substitution method starts with the structure of a known 2D material and replaces one or more element in this material with other elements either in the same group or its neighbor elements.

Both the element substitution method and the de novo generation method start with known 2D crystal structures. Currently, there are several open-source 2D material databases generated through exfoliation, substitution, or de novo generation methods. The Computational 2D Materials Database (C2DB)\cite{C2DB1,C2DB2} uses both exfoliation and substitution methods to organize a wealth of computed properties for 4038 (checked in October 2022) atomically thin 2D materials. The materials in the C2DB comprise both experimentally known and not previously synthesized structures. They have been generated in a systematic fashion by the combinatorial decoration of different 2D crystal lattices. Starting from 108,423 unique, experimentally known 3D compounds, MC2D\cite{MC2D} uses only the exfoliation method to identify a subset of 5,619 compounds that appear layered according to robust geometric and bonding criteria (checked in October 2022). High-throughput calculations using van der Waals density functional theory (DFT) validated against experimental structural data and calculating random phase approximation binding energies further allowed the identification of 1,825 compounds that can be either easily or potentially be exfoliated. 2D Materials Encyclopedia (2DMatPedia) database\cite{2DMatPedia,2DMatPedia2} screened all bulk materials in the database of Materials Project for layered structures by a topology-based detection algorithm and theoretically exfoliated them into monolayers. Then, new 2D materials are generated by applying chemical substitution of elements in known 2D materials by other elements from the same group in the periodic table. There are a total of 6351 materials in the current 2DMatPedia database (checked in December 2022), whereas 2940 were obtained by exfoliating existing layered materials (top-down approach), 3409 were obtained by the chemical substitution of 2D materials (bottom-up approach), and 2 were obtained from the literature via neither a top-down or bottom-up approach. The bottom-up approach starts from the 35 unary and 755 binary compounds obtained from the top-down approach. Only the same-column elements are used for substitution. By employing 22 different 2D crystal prototypes and 52 different chemical elements from the periodic table, Virtual 2D Materials Database (V2DB)\cite{V2DB} applied a brute-force substitution method to generate a systematic library of more than 72 million 2D compounds. Next, symmetry, neutrality, and stability sequential filtering layers are applied to identify 316,505 likely stable 2D materials.

Materials Cloud\cite{MatC} provides a practical yet straightforward approach to assessing whether any 3D compound can be exfoliated into 2D layers. The multistep procedure starts by pre-screening layered structures based on geometrical criteria requiring only the atomic positions of the atoms in the structure. The resulting filtered structures are featurized, and finally, an ML model based on a random forest classifier is applied to assess whether the material can be exfoliated or, instead, has high binding energy (HBE).
Friedrich et al.\cite{friedrich2022data} outlined a new set of non-vdW 2D materials by employing data-driven concepts and extensive calculations. By filtering the AFLOW-ICSD database according to the structural prototype of the two experimentally realized systems Fe$_2$O$_3$ and FeTiO$_3$, they have obtained 8 binary and 20 ternary 2D material candidates. The most recent approach for 2D material generation is based on the deep learning generative model. Lyngby et al.\cite{lyngby2022data} use a crystal diffusion variational
autoencoder (CDVAE)\cite{CDVAE} to generate new 2D structures of high chemical
and structural diversity and with formation energies mirroring the training structures. They also use the element substitution method to generate new possible 2D materials based on the newly generated 2D structures. In total, they find 11630 predicted new 2D materials, where 3073 are generated by CDVAE and 8,599 come from element substitution of these 3073 structures. They find that 2,004 of their generated 2D candidates are within 50 meV of the convex hull and could potentially be synthesised. In order to capture all these structural features of 2D and quasi-2D materials, Wang et al.\cite{wang2012effective} developed a new 2D structure search module in CALYPSO code that is based on 2D PSO algorithm but allows the relaxation of atomic coordinates in the perpendicular direction. They predicted a new family B$_x$N$_y$ with different chemical compositions that have layered structures.

Here we propose a computational pipeline, material transformer generator(MTG), for generative discovery of new 2D materials (and other crystal materials). Our method is based on combining a 2D composition generator trained with known 2D material compositions, two template based crystal structure predictors, two machine learning potential-based structure relaxers, and DFT relaxation. Extensive experiments show that our MTG pipeline can be used discover a large number of hypothetical 2D materials. 

\section{Method}

\begin{figure}[ht] 
    \centering
    \begin{minipage}[c]{1.0\textwidth}
        \centering
        \includegraphics[width=\textwidth]{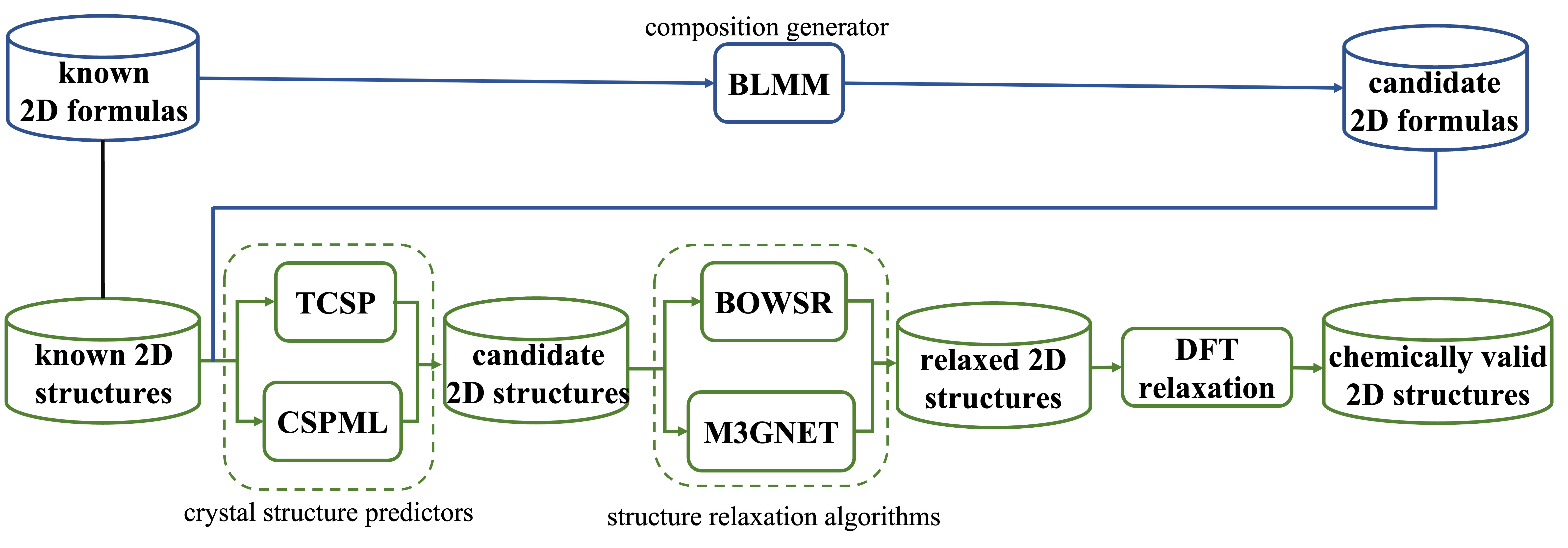}
    \end{minipage}\\
   
    \caption{Architecture of our material transformer generator(MTG) pipeline. BLMM is a tranformer neural network based composition generator. TCSP and CSPML are template based crystal structure prediction algorithms; and BOWSR and M3GNET are machine learning potential based structure relaxing algorithms. DFT relaxation is a first-principles calculation method.}
    \label{fig:flowchart}
\end{figure}

Figure\ref{fig:flowchart} (a) shows how the framework of our MTG pipeline for 2D material generation. We collected known 2D formulas and their structures from open datasets C2DB, MC2D, 2DMatPedia, and V2DB. We then train a set of BLMM (blank language models for materials) composition generators with known 2d formulas to generate new 2D formulas.
Next, we use known 2D structures as templates for structure prediction of these candidate 2D formulas using two crystal structure prediction algorithms TCSP and CSPML. TCSP is a template-based crystal structure prediction algorithm based on oxidation state patterns. CSPML is a machine learning-based crystal structure prediction method using a machine learning model to select templates. For a given new 2D formula such as SrTiO$_3$, both models will first select all template structures with prototype ABC3, but they are very different when sorting all these templates. TCSP calculates the element mover distance score and elements oxidation states, which focus on element distance. However, CSPML selects candidates using structural similarity. This structure similarity measure uses only the topological features of the atomic coordinates and does not use any information about the elemental composition. 
After choosing the appropriate templates and generating new 2D structures, we use two machine learning potential-based relaxation algorithms to optimize the structures. The first one is BOWSR 
which is a Bayesian optimization with symmetry relaxation. The second one is M3GNET, which uses materials graph neural networks with 3-body interactions as energy estimation model for structure relaxation. The BOWSR algorithm relaxes each structure by changing the independent lattice parameters and atomic coordinates to obtain lower potential energy. During relaxation, the M3GNET algorithm takes all atom coordinates and the 3X3 lattice matrix into consideration. The attributes of the bond, atom, and state are updated in order. For each attribute update, all previous attributes of these three parameters are considered.
After all these operations, for all generated 2D formulas, we obtain near-equilibrium relaxed structures. After the fast machine learning potential-based relaxation, we further apply the DFT-based relaxation procedure to optimize the structures. Finally, we calculate the formation energy and e-above-hull energy of top structures to evaluate the final performance.

\subsection{BLMM: Transformer based 2D material composition generation}

\begin{figure}[ht] 
    \centering
    \begin{minipage}[c]{0.8\textwidth}
        \centering
        \includegraphics[width=\textwidth]{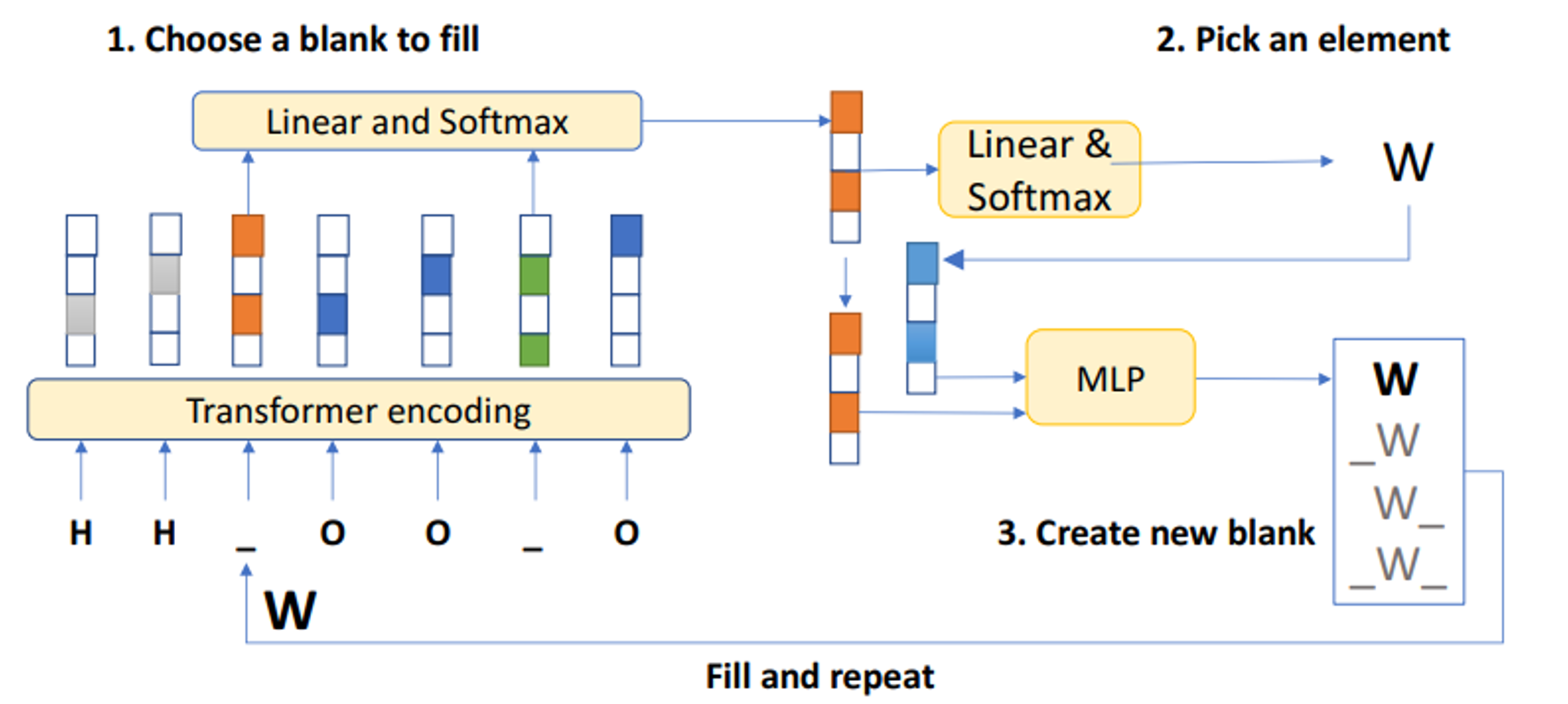}
    \end{minipage}\\
   
    \caption{BLMM architecture \cite{wei2022crystal}.}
    \label{fig:BLMM}
\end{figure}

The material composition can be mapped into a sequence generation problem as a composition such as SiTiO$_3$ can be conveniently expanded into a specific sequence (e.g. Si Ti O O O) sorted by the electronegativities of the elements. The BLMM model is a composition generator built on the latest transformer deep neural network models, shown to be excellent on sequence learning and sequence generation. 
By adopting the self-attention mechanism to weighting the significance of all tokens in the input sequence, the transformer model\cite{vaswani2017attention} has been proved as the state-of-the-art in the fields of natural language processing and computer vision. Based on the traditional transformer, Shen et.al.\cite{BLM} proposed a blank language model (BLM) which could generate sequences by dynamically creating and filling in blanks. 
Our BLMM composition generator\cite{wei2022crystal}  is developed based on the BLM blank-filling model.
All material formulas can be rewritten as sequences (e.g., SiTiO$_3$ to Si Ti O O O) composed of a vocabulary with 118 or fewer elements. We then train a BLMM based 2D composition generator using our 2D materials dataset. The architecture of the BLMM algorithm is shown in Figure\ref{fig:BLMM}. Generation starts with a single blank and ends when there is no blank. In each step, the model selects a blank, predicts a word w, and replaces the blank with the word w and possibly adjoining blanks. By repeating this blank selecting and filling process, a blank can be expanded to any number of words. 
Then we use this well-trained BLMM model to generate new 2D compositions. After getting the generated compositions, we first remove duplicate compositions that are already included in known 2D datasets and then take the nonredundant formulas as our new 2D material candidates to be fed to the step of template-based 2D material structure prediction.

\subsection{Template based 2D material structure prediction}

Currently, generic crystal structure prediction is still an unsolved problem despite that global optimization-based algorithms such as USPEX and CALYPSO can be applied to solve structures for small systems. On the other hand, we find that, similar to bulk materials\cite{mehl2017aflow,su2017construction,griesemer2021high}, most existing 2D material structures can be categorized into a very limited number of structure prototypes, which implies that their structures can be obtained using template-based elemental substitution. 

After composition generation and duplicate checking, we obtain a 2D material composition candidates dataset. To gain the probable structures of all candidates, we use two different template-based element substitution methods to select the most similar structure template and then use element substitution to generate target structures. The crystal structure generated by these two methods has the same lattice parameters and atomic coordinates as the template structure and needs to be further relaxed.

\begin{figure}[ht] 
    \centering
   
    \begin{minipage}[c]{0.45\textwidth}
        \centering
        \includegraphics[width=\textwidth]{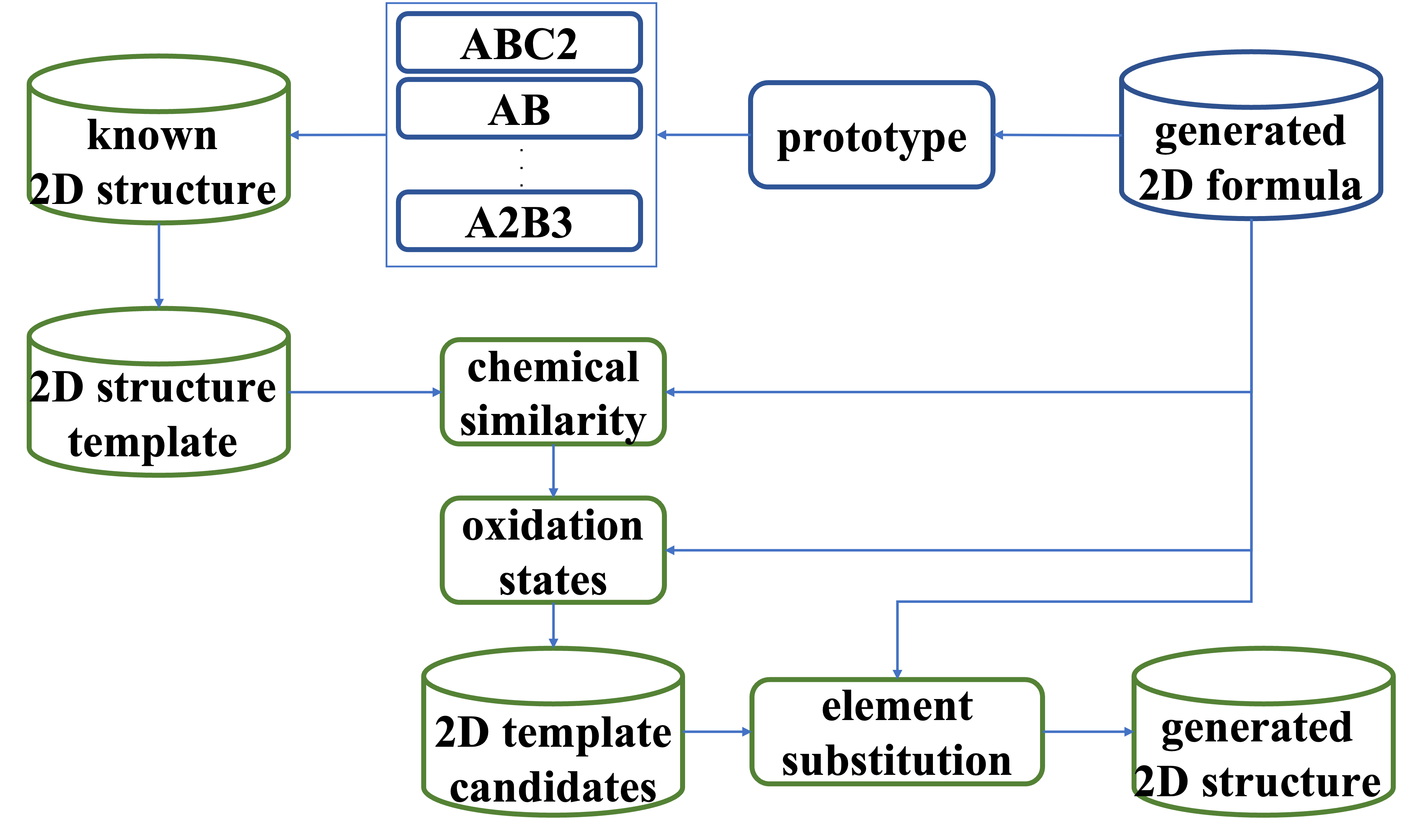}
        \subcaption{}
    \end{minipage}
    \begin{minipage}[c]{0.45\textwidth}
        \centering
        \includegraphics[width=\textwidth]{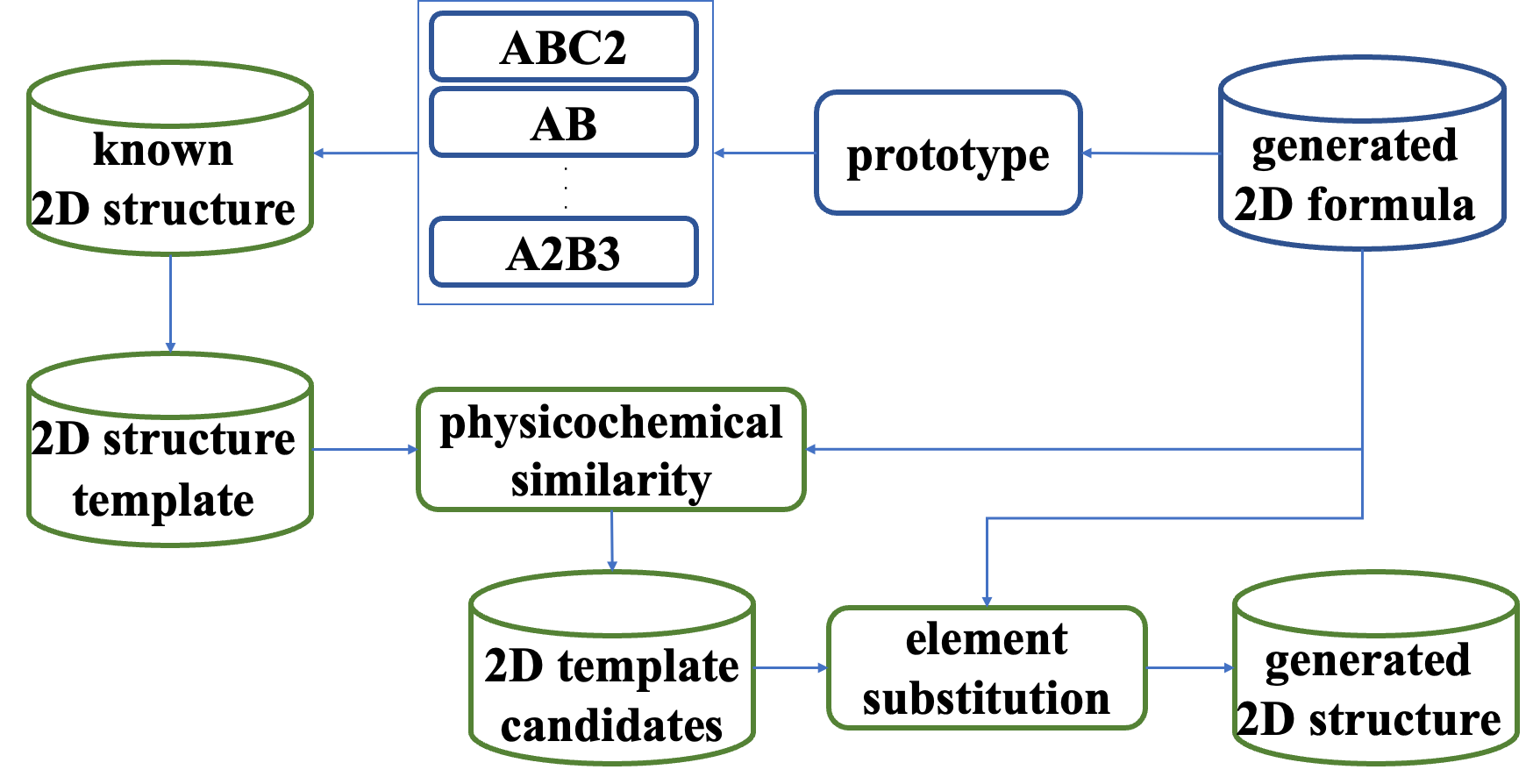}
        \subcaption{}
    \end{minipage}\\
   
    \caption{Template-based CSP algorithms. (a) TCSP architecture.\cite{wei2022tcsp}. (b) CSPML architecture\cite{CSPML}}
    \label{fig:tsp}
\end{figure}

\paragraph{TCSP: is a template-based crystal structure prediction algorithm.} 
The architecture of the TCSP algorithm is shown in Figure\ref{fig:tsp} (a). For a given candidate 2D material formula, the TCSP algorithm first searches all known 2D material structure templates that share the same composition prototype as this formula (e.g., SiTiO$_3$ has prototype ABC$_3$). The Element’s mover distance(ElMD)\cite{ElMD} is used to measure the compositional similarity between the query formula and compositions of all possible template structures. It then picks the top 5 structures with the smallest compositional distances as the candidate templates . For each of these candidate templates, we use the Pymatgen\cite{Pymatgen} package to check whether it has the same oxidation states as the query formula. Templates with identical oxidation states are then added to the final template list. If no such templates are found, all five of the top structures are taken as the final templates. Pymatgen’s StructureMatcher module is then used to reduce the redundant template structures. For each structure cluster, only one of these structures can be kept, which can significantly reduce the number of similar structure templates. After the templates of the query formula are designated, the algorithm then enumerates all of the possible element substitution pairs between the query and template formulas as Algorithm A2 in Ref\cite{wei2022tcsp}. It is possible that there are several element pair substitutions in one template to get the target formula. A replacement quality score is then calculated by summing the ElMD of all element pairs' substitution arrangements. This score represents how similar the substitution element pairs are. A lower score means higher similarity and thus higher quality.

\paragraph{CSPML is a machine learning-based crystal structure prediction algorithm.} 
CSPML relies on metric learning\cite{Metric} for crystal structure prediction, which can select template structures from known structure databases with high similarity to the given composition. Metric learning uses a binary classifier to distinguish whether two given compositions have similar structures as defined by a similarity threshold of local structure order parameters (LoStOPs)\cite{LoStOPs}. The architecture of the CSPML algorithm is shown in Figure\ref{fig:tsp} (b). For a given 2D formula, CSPML first restricts the candidates to structures with the same compositional ratio (e.g., SiTiO$_3$ has a composition ratio of 1:1:3). The compositional descriptor of query formula and templates is then calculated by XenonPy\cite{XenonPy}. XenonPy provides 58 physicochemical features for each element. For a given composition, by calculating the weighted mean, weighted sum, weighted variance, min-pooling, and max-pooling of all elements, XenonPy generates a 290-dimensional (58x5) descriptor vector. A traditional multi-layer perceptron (MLP) is used to figure out how similar the template structure and the query formula are. The absolute difference between two compositional descriptors is used as the input. We pick the top five template structures with the biggest similarity scores with the query formula as the template candidates. The structure of the query formula is then generated by replacing the atoms in the templates with atoms in the query composition. When two or more elements have the same composition ratio, the substitution element pairs are not uniquely determined. In such cases, we substitute a pair of elements with the most similar physicochemical properties, as described in Ref\cite{CSPML}.

\subsection{Structure relaxation}

Accurately predicting novel stable crystal structures and their properties is a fundamental goal in computation-guided materials discovery. While ab initio approaches such as density functional theory (DFT) have been phenomenally successful in this regard, their high computational cost and poor scalability have limited their broad application across the vast chemical and structural spaces. 
To circumvent this limitation, machine learning has emerged as a new paradigm for developing efficient surrogate models for predicting material properties at scale. In this paper, after gaining basic structures from template-based element substitution methods, we apply and compare two different machine-learning potential-based structure relaxation methods.

\paragraph{BOWSR: Bayesian optimization with symmetry relaxation algorithm}
BOWSR is a graph-based neural network-based structure relaxation algorithm that uses Bayesian optimization as an optimizer. Bayesian optimization is an adaptive strategy for the global optimization of functions. In the crystal structure relaxation problem, the target function that needs to be optimized is the potential energy surface, which describes the energy of the crystal structure. During the relaxation process of the BOWSR algorithm, the symmetry of the lattice and the Wyckoff positions of the atoms are limited. Only the independent lattice parameters and atomic coordinates are allowed to change. The BOWSR algorithm sets parameters for each structure based on these changeable, independent lattice parameters and atomic coordinates. The potential energy surface of each training observation is calculated by a graph neural network (MEGNet) energy model, which is trained with 12,277 stable structures with DFT-calculated formation energies. Bayesian optimization is then used to relax structures iteratively towards states with lower energies.

The geometry relaxation of a structure of N atoms requires optimizing 3N + 6 variables, 3 fractional coordinates for each atom, and 6 lattice parameters in total. By keeping the symmetry the same during the relaxation process, it can reduce the number of independent variables. New structures are generated by Bayesian optimization, which minimizes the formation energy, and the changed variables are then used as inputs to predict new energy. The previous step is then repeated multiple times until the formation energy reaches the lowest point or reaches the maximum number of iterations. The final structure is the BOWSR relaxation results.

\paragraph{M3GNET: materials graph neural networks with 3-body interactions}

M3GNet (M3GNET) is a new materials graph neural network architecture that incorporates 3-body interactions for formation energy prediction. It combines graph-based deep learning interatomic potential (IAP) and many-body features of traditional IAPs with those of flexible graph material representations. The inputs of the M3GNet model are position-included graphs. The atomic numbers of elements and the pair bond distance in the input graph are embedded as graph features. The three-body and many-body interaction atom indices and angles are calculated by the many-body computation module. The bond and atom information is then updated through a graph convolution module. 

A key difference with prior materials graph implementations such as MEGNet is the addition of the coordinates for atoms and the 3x3 lattice matrix in crystals, which are necessary for obtaining tensorial quantities such as forces and stresses via auto-differentiation. The difference of M3GNet with BOWSR's GNN potential is that it is trained with both stable structures and unstable structures. The M3GNet-based relaxation algorithm \cite{M3GNET} is also different from BOWSR's Bayesian optimization. It uses an algorithm named FIRE, which is derived from molecular dynamics with additional velocity modifications and adaptive time steps and inertia to achieve fast inertial relaxation engine.
The ability of a M3GNet-based relaxing algorithm to accurately and rapidly relax arbitrary crystal structures and predict their energies makes it ideal for large-scale materials discovery.

\subsection{DFT calculations}

 We carried out the first-principles calculations based on the density functional theory (DFT) using the Vienna \textit{ab initio} simulation package (VASP)\cite{Vasp1,Vasp2,Vasp3,Vasp4} to optimize the candidate structures suggested by the machine learning models. The projected augmented wave (PAW) pseudopotentials were used to treat the electron-ion interactions\cite{PAW1, PAW2} with 520 eV plane-wave cutoff energy.  The generalized gradient approximation (GGA) based Perdew-Burke-Ernzerhof (PBE) method was considered for the exchange-correlation functions \cite{GGA1, GGA2}. The energy convergence criterion was  10$^{-5}$ eV and  the force convergence criterion was 10$^{-2}$ eV/{\AA} for all the DFT calculations. The Brillouin zone integration for the unit cells was performed  employing the $\Gamma$-centered  Monkhorst-Pack $k$-meshes. The formation energies (in eV/atom) of the materials were determined employing the formula in  Eq.~\ref{eq:form}, where $E[\mathrm{Material}]$ is the total energy per unit formula of the target structure, $E[\textrm{A}_i]$ is the energy of $i^\mathrm{th}$ element of the material, $x_i$ indicates the number of A$_i$ atoms in a unit formula, and $n$ is the total number of atoms in a unit formula($n=\sum_i x_i$).  The Pymatgen code\cite{Pymatgen} was used to compute the energy above hull values of the materials with negative formation energies.

\begin{equation}
    E_{\mathrm{form}} =\frac{1}{n}(E[\mathrm{Material}] - \sum_i x_i E[\textrm{A}_i])
    \label{eq:form}
\end{equation}

\subsection{Evaluation criteria}

We use a series of performance metrics to evaluate our 2D material generation pipeline. 
To evaluate the BLMM 2D material composition generator, we calculate the validity, uniqueness, recovery rate, and novelty. Formation energy is used as an indicator to evaluate the template-based 2D structure generator and relaxer. To further verify the structures, we use VASP to calculate the energy-above-the-hull.   

\textbf{Validity.} For all formulas generated by the BLMM algorithm, we use Semiconducting Materials by Analogy and
Chemical Theory(SMACT)\cite{davies2019smact} to check whether they obey the charge neutrality and electronegativity (CNEN) rules. 

\textbf{Uniqueness.} Uniqueness percentage is calculated by using the number of unique samples divided by the total generated samples. The uniqueness indicator shows the BLMM model's ability to generate diverse samples.

\textbf{Recovery Rate and Novelty.} To check the BLMM model's capability to generate novel materials, we calculate the recovery rate and novelty of generated formulas. The recovery rate shows the percentage of training samples that have been rediscovered. Novelty shows how many new samples have been generated.

\textbf{Formation energy.} The way to evaluate the structure generation models is to check the stability of generated structures. For structures generated and relaxed through our pipeline, we calculate their formation energy using M3GNET. 

\textbf{Energy above convex hull.} The energy convex hull \cite{liu2015spinel} is generated based on existing stable structures. Structures with energy lying on the convex hull are thermodynamically stable, and the ones above it are either metastable or unstable. For all structures with negative formation energy, we use the energy above convex hull as a further filter to select more stable structures.

\subsection{Hyperparameters and training}

For 2D formula generation, each BLMM model trained on 2D datasets generates 100,000 samples. After generation, we use the TCSP and CSPML methods separately to generate structure candidates for these samples. BOWSR and M3GNET are then used to relax generated structures. Table\ref{table:hyperparameter} shows the hyperparameters used in the BLMM, TCSP, CSPML, BOWSR, and M3GNET models. In the BLMM model, we use an element vocabulary with size 130, and the generated formula sequence length is limited to 205. The maximum number of tokens per batch is set to 40,000, and the number of training steps is set to 200,000.
The candidate template structure number of both the TCSP and CSPML models is set to 10. And only the top 5 candidates, sorted by ElMD and XenenPy, respectively, can be used as real templates. Relax method BOWSR uses a Bayesian optimizer with initial points 1000, iteration steps 1000, and seed number 42. The M3GNET relax method uses FIRE\cite{FIRE} optimizer with a 0.1 total force tolerance for relaxation convergence and 500 relax steps.

\begin{table}[h]
\begin{center}
\caption{Hyperparameters used in models.}
\label{table:hyperparameter}
\begin{tabular}{|cc|cc|cc}
\hline
\multicolumn{2}{|c|}{TCSP}                                 & \multicolumn{2}{c|}{CSPML}                     & \multicolumn{2}{c|}{BLMM}                                       \\ \hline
\multicolumn{1}{|c|}{candidate}       & 10                 & \multicolumn{1}{c|}{candidate}       & 10      & \multicolumn{1}{c|}{data workers} & \multicolumn{1}{c|}{32}     \\ \hline
\multicolumn{1}{|c|}{top}             & 5                  & \multicolumn{1}{c|}{top}             & 5       & \multicolumn{1}{c|}{max steps}    & \multicolumn{1}{c|}{200000} \\ \hline
\multicolumn{1}{|c|}{sort}            & ElMD               & \multicolumn{1}{c|}{sort}            & XenonPy & \multicolumn{1}{c|}{max token}    & \multicolumn{1}{c|}{40000}  \\ \hline
\multicolumn{1}{|c|}{filter}          & ratio              & \multicolumn{1}{c|}{filter}          & ratio   & \multicolumn{1}{c|}{vocab size}   & \multicolumn{1}{c|}{130}    \\ \hline
\multicolumn{1}{|c|}{filter}          & oxidation          & \multicolumn{1}{c|}{}                &         & \multicolumn{1}{c|}{max len}      & \multicolumn{1}{c|}{205}    \\ \hline
\multicolumn{2}{|c|}{BOWSR}                                & \multicolumn{2}{c|}{M3GNET}                    &                                   &                             \\ \cline{1-4}
\multicolumn{1}{|c|}{optimizer}       & Bayesian Optimizer & \multicolumn{1}{c|}{optimizer}       & FIRE    &                                   &                             \\ \cline{1-4}
\multicolumn{1}{|c|}{initial points}  & 100                & \multicolumn{1}{c|}{force tolerance} & 0.1     &                                   &                             \\ \cline{1-4}
\multicolumn{1}{|c|}{iteration steps} & 100                & \multicolumn{1}{c|}{relax steps}     & 500     &                                   &                             \\ \cline{1-4}
\multicolumn{1}{|c|}{seed}            & 42                 & \multicolumn{1}{c|}{}                &         &                                   &                             \\ \cline{1-4}
\end{tabular}
\end{center}
\end{table}

\section{Results}

\subsection{Datasets}

As shown in Table\ref{table:dataset}, our template-based 2D materials generation models are trained using the materials downloaded from the C2DB\cite{C2DB1,C2DB2}, MC2D\cite{MC2D}, 2DMatPedia\cite{2DMatPedia,2DMatPedia2}, and V2DB\cite{V2DB} databases with a total of 328,719 formula samples and 12,214 structures. The C2DB dataset was initially generated by decorating an experimentally known crystal structure prototype with atoms chosen from a (chemically reasonable) subset of the periodic table. The MC2D dataset starts from experimentally known 3D compounds and finds 1,825 compounds that are either easily or potentially exfoliate. The 2DMatPedia dataset is searched from the Materials Project database\cite{MP} and uses exfoliate first to find possible 2D structures, and new structures generated by exfoliation are then used as templates of element substitution. The generation of the V2DB dataset employs the brute-force element substitution method. This method generated 72,522,240 possible combinations of 2D materials and only 0.4\% of these passed the symmetry, neutrality, and stability validation.

In this work, we separated these 2D samples into two datasets. The first one is an experimental dataset (exp2d for short) with 4,023 formulas and corresponding structures from the C2DB and MC2D datasets.  The second dataset (all2d for short) contains the all the samples in the above-mentioned our known 2D databases with a total of 302,174 unique formulas and 8,019 structures.

\begin{table}[h]
\begin{center}
\caption{Open source datasets used in 2D material discovery.}
\label{table:dataset}
\resizebox{\columnwidth}{!}{\begin{tabular}{|c|c|c|c|c|cc|}
\hline
Dataset    & Formula & Structure & Type         & From                                                         & \multicolumn{1}{c|}{Exfoliation} & Substitution \\ \hline
C2DB       & 4038    & 4038      & experimental & known 2D crystal structure prototype & \multicolumn{1}{c|}{N/A}   &  all                  \\ \hline
MC2D       & 1825    & 1825      & experimental & known 3D compounds              & \multicolumn{1}{c|}{all}         & N/A          \\ \hline
2DMatPedia & 6351    & 6351      & calculate    & Material Project                                                           & \multicolumn{1}{c|}{2940}        & 3409         \\ \hline
V2DB & 316505    & N/A      & calculate    & 22 known 2D crystal prototypes                                                         & \multicolumn{1}{c|}{N/A}        & all        \\ \hline
\end{tabular}}
\end{center}
\end{table}

\subsection{Composition generation performance}

We use the BLMM algorithm to generate new 2D material compositions based on two different datasets, the experimental 2D dataset, exp2d, and an all 2D dataset, all2d. For each dataset, we train a generation model using these formulas and then use these well-trained models to generate new formulas. We also use transfer learning to train BLMM models on the materials project database and then fine-tune these pre-trained models using our two datasets, which are named all2d-transfer and exp2d-transfer, respectively. Four composition generation models are trained to generate 100,000 formulas separately. The generated results are shown in Table\ref{table:composition}. Furthermore, to check whether generated formulas are chemically valid, we employ two filters to check their charge neutrality (CN) and electronegativity (EN), this checking step is called CNEN for short. The results are shown in Table\ref{table:composition}, 93.3\%, 67.1\%, 93.2\%, and 67.0\% generated compositions passed the CNEN check. Besides, we remove duplicate composition in each model, and they achieve 65.8\%, 26.5\%, 69.4\%, and 21.0\% uniqueness respectively. We also calculate the recovery rate and novelty of these generational models. As we can see in the Table \ref{table:composition}, their recovery rates are  0.6\%, 9.2\%, 0.6\%, and 11.2\% while the novelties are 63.7\%, 24.6\%, 67.5\%, and 18.8\%. This evaluation demonstrates that our methods have the ability to generate stable and innovative compositions that form stable 2D structures. 
Since the exp2d dataset is smaller than the all2d dataset, the BLMM model trained with the exp2d dataset has much fewer samples to learn from and use for interpolation. Thus, the BLMM model trained with the exp2d dataset has lower validity, uniqueness, and novelty percentages than the BLMM model trained with the all2d dataset. However, the recovery rate of the BLMM model trained with fewer samples is higher because the interpolation space is smaller and the interpolated results generated by the BLMM model are more likely to be the same as the training samples.
The composition generator pre-trained with the materials project database and fine-tuned with the all2d dataset achieves higher uniqueness and novelty compared with the generator only trained with the all2d dataset. However, due to the lack of sufficient transfer learning samples, the BLMM model fine-tuned with the exp2d dataset has lower uniqueness and novelty compared with the BLMM model trained with the exp2d dataset.

\begin{table}[h]
\begin{center}
\caption{Composition generation results.}
\label{table:composition}
\begin{tabular}{|c|c|c|c|c|}
\hline
                  & all2d  & exp2d  & all2d-transfer & exp2d-transfer \\ \hline
Generated results & 100000 & 100000 & 100000 & 100000         \\ \hline
Validity          & 93.3\%  & 67.1\%  & 93.2\%   & 67.0\%          \\ \hline
Uniqueness        & 65.8\%  & 26.5\%  & 69.4\%   & 21.0\%          \\ \hline
Recover rate      & 0.6\%   & 9.2\%   & 0.6\%    & 11.2\%          \\ \hline
Novelty           & 63.7\%  & 24.6\%  & 67.5\%   & 18.8\%          \\ \hline
\end{tabular}
\end{center}
\end{table}

\FloatBarrier

\subsection{Distribution of generated candidate 2D compositions}

\begin{figure}[ht] 
    \centering
    \begin{minipage}[c]{0.45\textwidth}
        \centering
        \includegraphics[width=\textwidth]{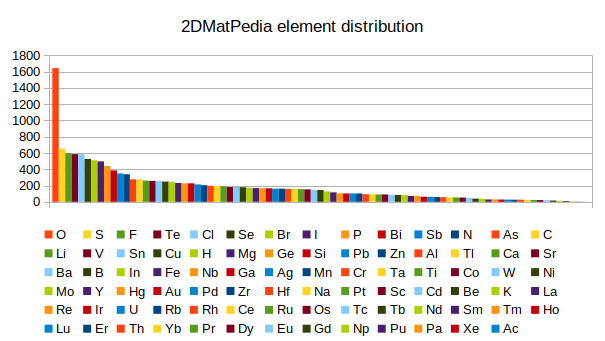}
        \subcaption{}
    \end{minipage}
    \begin{minipage}[c]{0.45\textwidth}
        \centering
        \includegraphics[width=\textwidth]{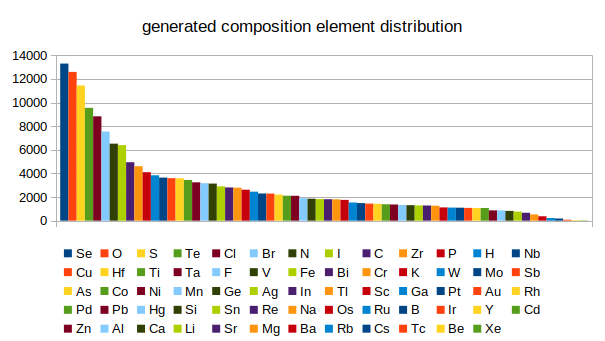}
        \subcaption{}
    \end{minipage}
     \caption{Elements distribution in training and generating samples. (a)Element distribution in the 2DMatPedia dataset. (b)Element distribution in BLMM model generated samples.}
    \label{fig:element distribution}
\end{figure}

To check the composition generation performance of BLMM, we plot the element distribution of compositions in the 2DMatPedia dataset and our generated samples in Figure\ref{fig:element distribution}, where (a) and (b) show the element frequency in compositions in the 2DMatPedia dataset and the BLMM model generated dataset, respectively. Here we take the BLMM model trained with the exp2d dataset as an example. The top 5 most frequent elements in the 2DMatpedia dataset are O, S, F, Te, and Cl. Out of the total 6,351 formulas, element O has appeared 1,642 times, or about 26\% of the 2Dmatpedia dataset. The occurrences of the elements S, F, Te, and Cl are 653, 598, 586, and 584, respectively. The top 5 most frequent elements in our generated dataset are Se, O, S, Te, and Cl. The element Se has shown 13,294 times out of the whole set of 67,103 formulas, which is about 20\% of the whole generated dataset. The elements O, S, Te, and Cl appear 12,591, 11,440, 9,546, and 8,829 times respectively. Overall, we find that four out of the top five most common elements in these two datasets are the same, and six in the top ten most common elements are also the same, indicating that the BLMM generator has learned the key composition preferences of 2D materials.

\begin{figure}[h] 
    \centering
    \begin{minipage}[c]{0.45\textwidth}
        \centering
        \includegraphics[width=\textwidth]{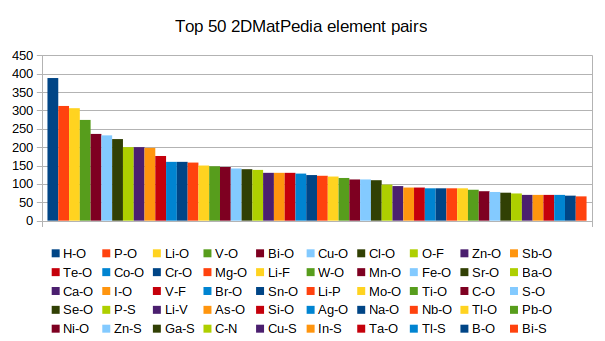}
        \subcaption{}
    \end{minipage}
    \begin{minipage}[c]{0.45\textwidth}
        \centering
        \includegraphics[width=\textwidth]{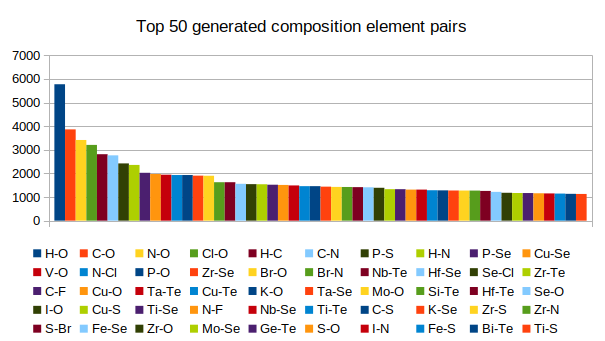}
        \subcaption{}
    \end{minipage}
     \caption{Top 50 element pairs distribution in training and generating samples. (a)Element pairs distribution in 2DMatPedia dataset. (b) Element pairs distribution in generated samples by BLMM model trained with exp2d dataset.}
     \label{fig:element pair}
\end{figure}

We also analyze the distribution of element pairs in the known 2D dataset and our generation results. To count the frequencies of element pairs, we take each of the possible 2-element combinations from the element set and count the number of compositions that contain this pair (we ignore the order of the two elements in the pair). The distribution of the top 50 element pairs in the 2DMatPedia and our generation datasets are shown in Figure\ref{fig:element pair} (a) and (b), respectively.  
The top 5 most frequent element pairs in the 2DMatpedia dataset are H-O, P-O, Li-O, V-O, and Bi-O. However, only the H-O element pair is shown in the top 5 of our generation results. The other 4 element pairs in our generation results are C-O, N-O, Cl-O, and H-C. These two datasets only share 2 common element pairs in the top 10 most frequent ones.

\begin{figure}[h] 
    \centering
    \begin{minipage}[c]{0.4\textwidth}
        \centering
        \includegraphics[width=\textwidth]{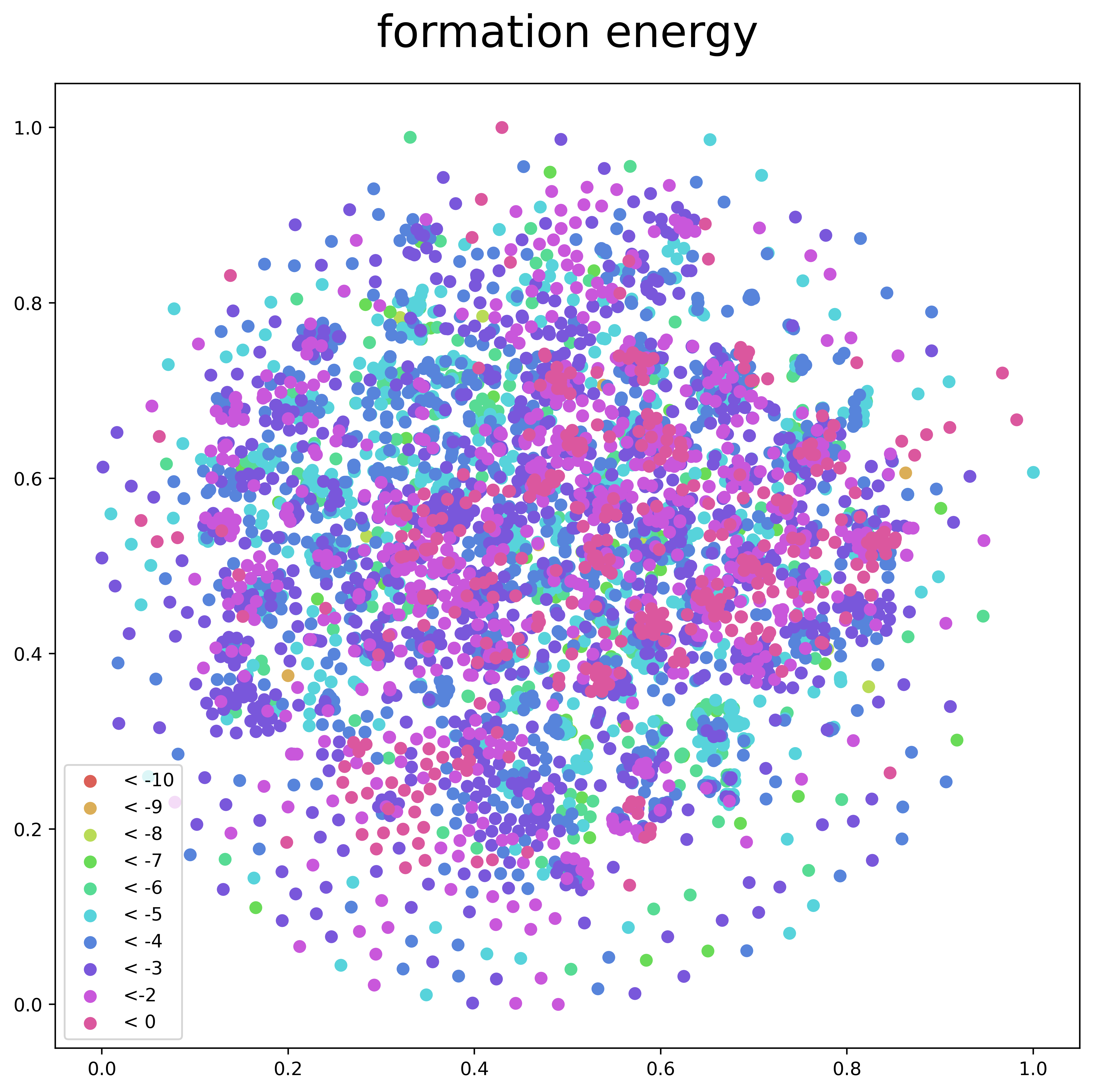}
        \subcaption{}
    \end{minipage}
    \begin{minipage}[c]{0.4\textwidth}
        \centering
        \includegraphics[width=\textwidth]{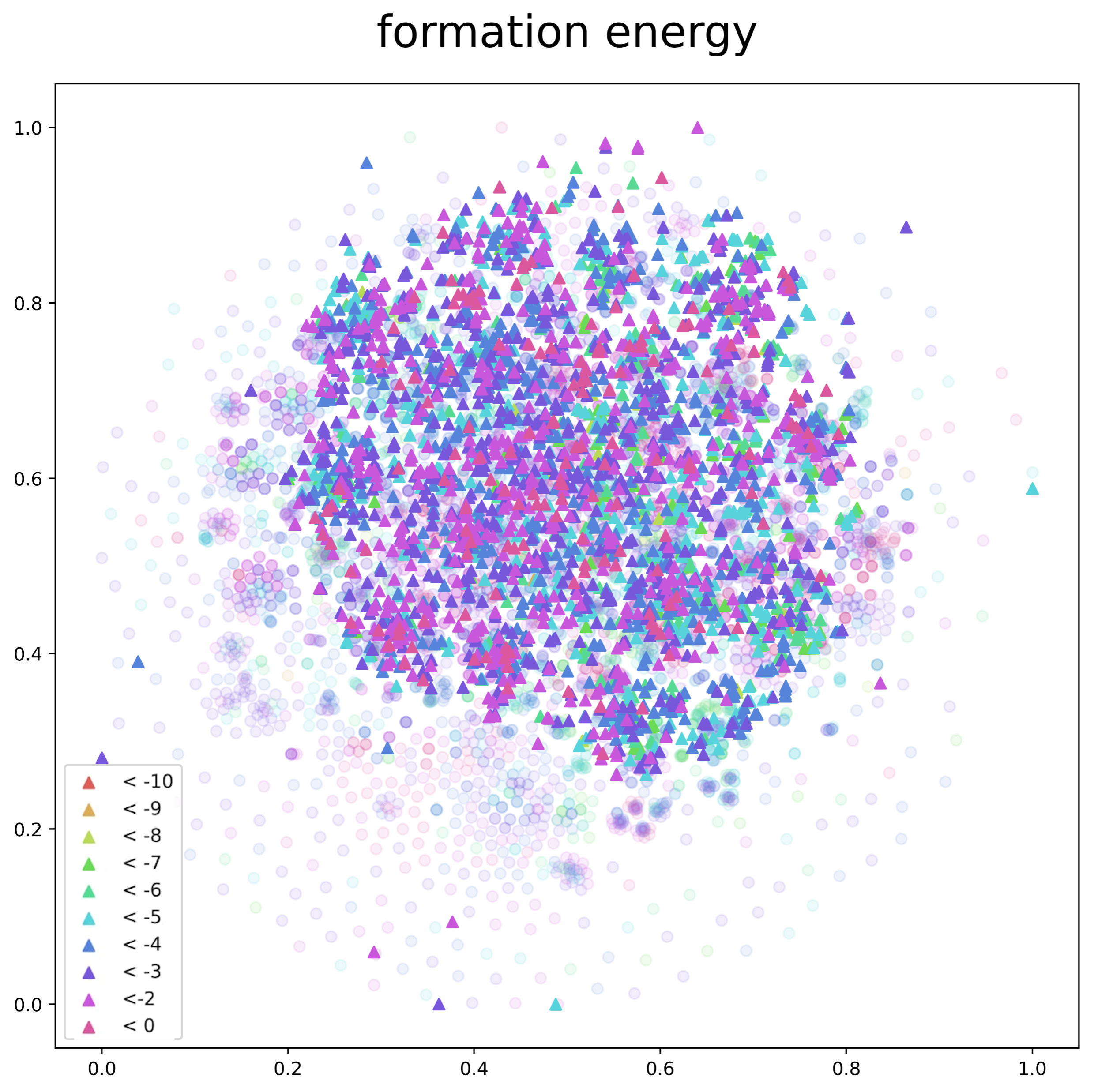}
        \subcaption{}
    \end{minipage}
     \caption{Formation energy distributions in the training (2DMatPedia) and generated samples(BLMM-exp2d). (a) Formation energy distribution in the 2DMatPedia dataset. (b) Formation energy distribution in generated samples.}
    \label{fig:tsne}
\end{figure}

To verify whether our newly generated compositions share a similar distribution with the known 2D compositions, we use the t-distributed Stochastic Neighbor Embedding(t-SNE)\cite{JMLR:v9:vandermaaten08a} technique to map the one-hot matrix of compositions to their corresponding formation energy. Each point in Figure\ref{fig:tsne} corresponds to one formula and the colors represent the formation energy levels. Figure\ref{fig:tsne} (a) shows the formation energy distribution of the 2DMatPedia samples. It can be found that most samples have formation energy between 0 and -3 eV/atom. Figure\ref{fig:tsne} (b) displays the formation energy distribution of our generated compositions by BLMM-exp2d, which are developed through the following pipeline: firstly, we train the BLMM model using the exp2d dataset and generate compositions; next, we generate candidate structures using the TCSP method and then relax these structures using the M3GNET model; thirdly, the formation energies of these structures are predicted by the M3GNET method. As the BLMM model is trained by adding and filling in blanks in existing materials, it has a strong interpolation capability when generating new samples. Therefore, the newly generated samples are always located around known samples.

\subsection{Stability distribution of generated samples}

Another way to check the quality of samples generated by our pipeline is to measure their formation energies and compare their distribution to that of the training set. We use formation energies predicted by the ML potentials M3GNET of both training samples and generated results.

\begin{figure}[h] 
    \centering
    \begin{minipage}[c]{0.45\textwidth}
        \centering
        \includegraphics[width=\textwidth]{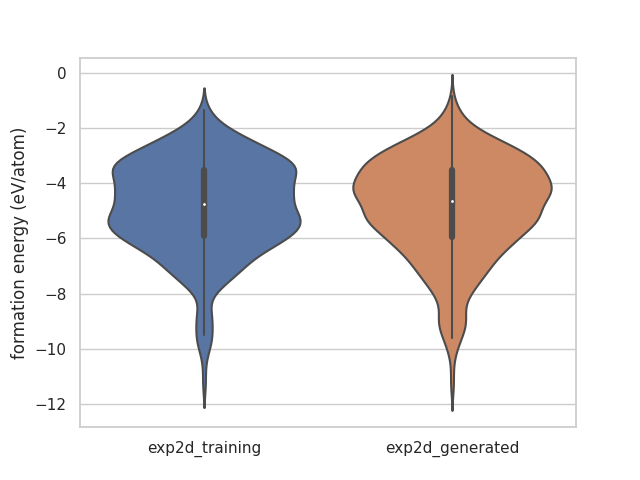}
        \subcaption{}
    \end{minipage}
    \begin{minipage}[c]{0.45\textwidth}
        \centering
        \includegraphics[width=\textwidth]{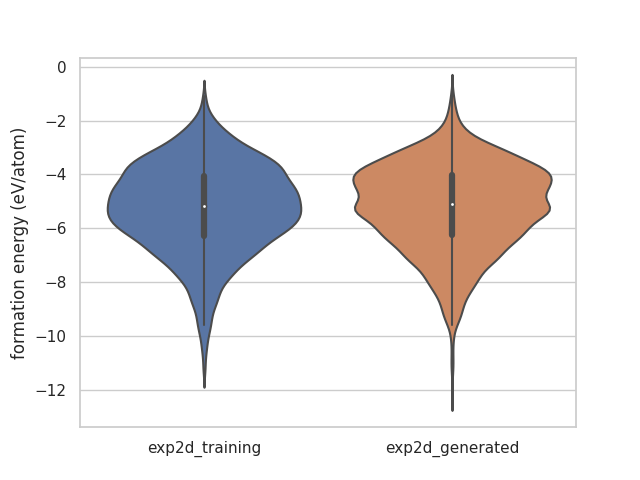}
        \subcaption{}
    \end{minipage}\\
    \begin{minipage}[c]{0.45\textwidth}
        \centering
        \includegraphics[width=\textwidth]{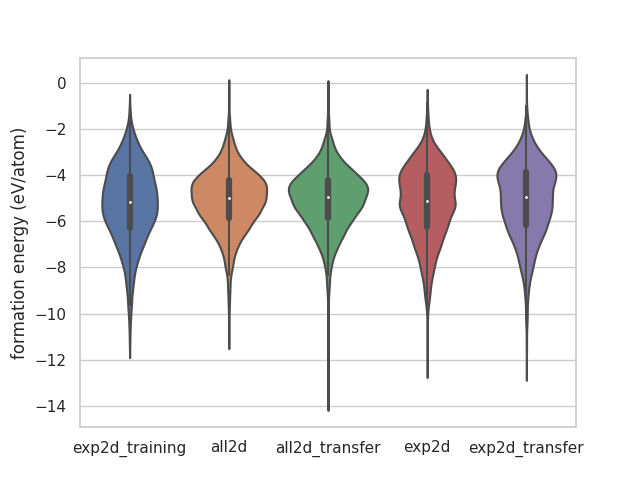}
        \subcaption{}
    \end{minipage}\\
     \caption{Formation energy per atom distribution. (a) Formation energy distribution of AB$_2$ type structures in the exp2d dataset and generated through our MTG pipeline (predicted by M3GNET). (b) Formation energy distribution of structures in the exp2d dataset and structures generated by our MTG-exp2d pipeline (formation energy predicted by M3GNET). (c) Formation energy distribution of compositions in the exp2d dataset and structures generated by our MTG pipelines (BLMM models trained by all four datasets, formation energies predicted by M3GNET).}
    \label{fig:formation energy distribution}
\end{figure}

We first check the formation energy distribution of a special material family AB$_2$, which is the most frequent prototype in all existing 2D datasets: C2DB, MC2D, and 2DMatPedia. There are 1,288 AB$_2$ samples in the exp2d dataset, and 1,928 AB$_2$ samples in our generated structures. The distributions of energies of these two datasets are shown in Figure\ref{fig:formation energy distribution}(a). 

Next, we check the structure-based formation energy distribution of the whole exp2d dataset samples and compare it with those of our generated samples. Figure\ref{fig:formation energy distribution}(b) shows that these two set of structures have similar formation energy, which means that new structures generated through our MTG pipeline are of high quality. 

Figure\ref{fig:formation energy distribution}(c) compares the energy distribution of samples in the exp2d training set with samples generated through out MTG pipeline. These compositions are generated by the BLMM model trained with four different datasets, as introduced in section 3.1. The energy distributions of formulas generated by BLMM trained with the all2d dataset and trained with the MP dataset but finetuned using the all2d dataset are very similar to each other. Same situation in BLMM models trained and finetuned with the exp2d dataset.

\FloatBarrier

\subsection{Discovery results}

Our MTG pipeline generates 148,563 candidate 2D formulas. For each formula, we generate 10 structures using TCSP and CSPML and then we do M3GNET based structure relaxation and we pick the top 1 structure with the lowest energy. Then we conduct DFT-based relaxation to generate final structures. 

\begin{figure}[ht] 
    \centering
    \begin{minipage}[c]{0.3\textwidth}
        \centering
        \includegraphics[width=\textwidth]{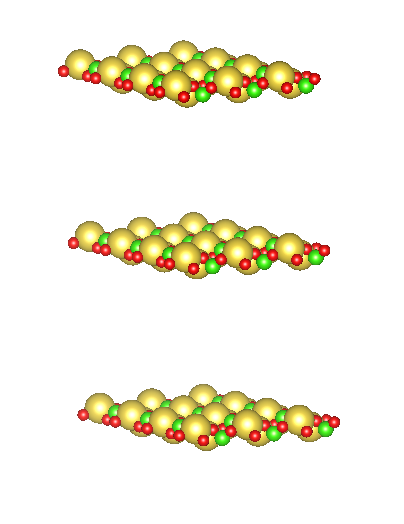}
        \subcaption{Template structure Na$_4$Ti$_2$S$_4$O$_2$}
    \end{minipage}
    \begin{minipage}[c]{0.3\textwidth}
        \centering
        \includegraphics[width=\textwidth]{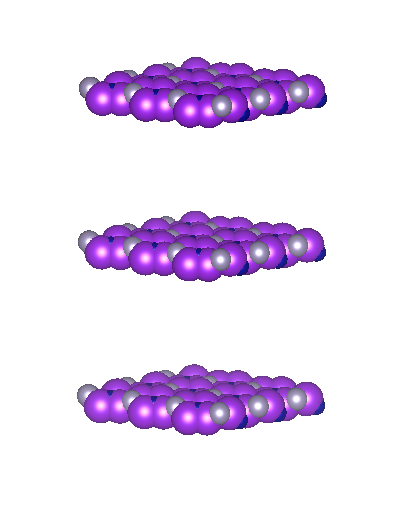}
        \subcaption{Generated structure K$_4$Cr$_2$Ge$_4$Te$_2$}
    \end{minipage}
    \begin{minipage}[c]{0.3\textwidth}
        \centering
        \includegraphics[width=\textwidth]{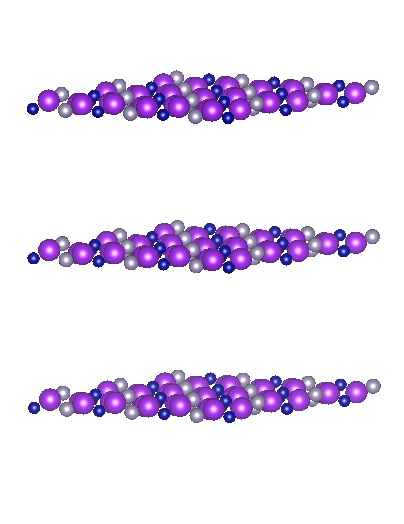}
        \subcaption{Relaxed structure K$_4$Cr$_2$Ge$_4$Te$_2$}
    \end{minipage}\\
    \begin{minipage}[c]{0.3\textwidth}
        \centering
        \includegraphics[width=\textwidth]{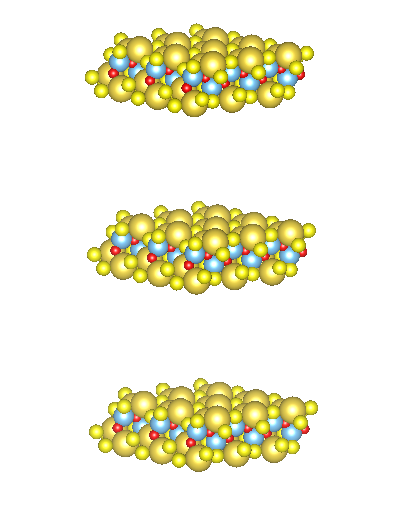}
        \subcaption{Template structure Na$_2$Cl$_2$O$_4$}
    \end{minipage}
    \begin{minipage}[c]{0.3\textwidth}
        \centering
        \includegraphics[width=\textwidth]{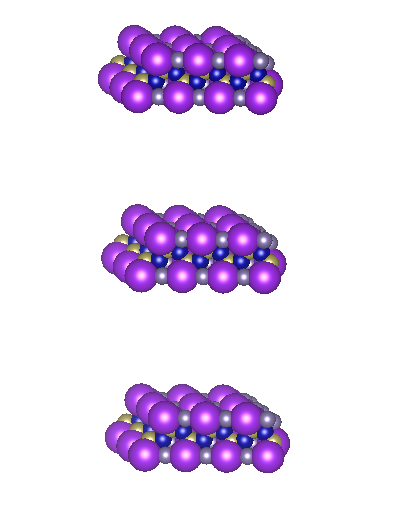}
        \subcaption{Generated structure K$_4$Cr$_2$Sn$_4$}
    \end{minipage}
    \begin{minipage}[c]{0.3\textwidth}
        \centering
        \includegraphics[width=\textwidth]{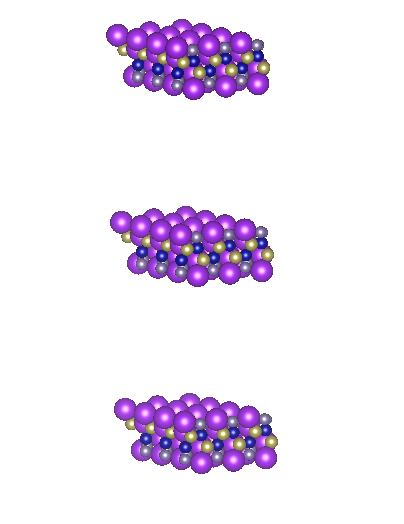}
        \subcaption{Relaxed structure K$_4$Cr$_2$Sn$_4$}
    \end{minipage}\\
    \caption{Illustration of the structure generation and relaxation process of our MTG pipeline. (a) to (c) shows the template structure, structure upon element substitution, and the fine-tuned structure after ML potential based relaxation for predicting the structure of K$_4$Cr$_2$Ge$_4$Te$_2$. (d) to (f) shows the similar process for K$_4$Cr$_2$Sn$_4$.}
    \label{fig:newstructure}
\end{figure}

Figure\ref{fig:newstructure} shows how we generated new structures based on specific template structures and how to relax newly generated structures to make them more stable. For formula K$_4$Cr$_2$Ge$_4$Te$_2$ generated by the BLMM algorithm, we first select the structure templates for predicting its crystal structure. As shown in Figure\ref{fig:newstructure} (a), TCSP picked Na$_4$Ti$_2$S$_4$O$_2$, a layered material, as the template structure. Figure\ref{fig:newstructure} (b) is then created through the TCSP algorithm. After relaxing by M3GNET, we get a more stable structure as shown in Figure\ref{fig:newstructure} (c). This relaxed structure is then sent to VASP to do further DFT relaxation and energy calculations. We can find that Figure\ref{fig:newstructure} (a) and (b) are more similar as in (b) only elemental substitutions are applied in the structure (a) with no atomic coordinate fine-tuning. Figure\ref{fig:newstructure} (c) changed the coordinates based on atom sizes and bond types to make this structure more structurally stable.
A similar procedure is applied to discover the structure of K$_4$Cr$_2$Sn$_4$, via the three steps as shown in Figure \ref{fig:newstructure} (d,e,f).

\begin{figure}[ht] 
    \centering
    \begin{minipage}[c]{0.24\textwidth}
        \centering
        \includegraphics[width=\textwidth]{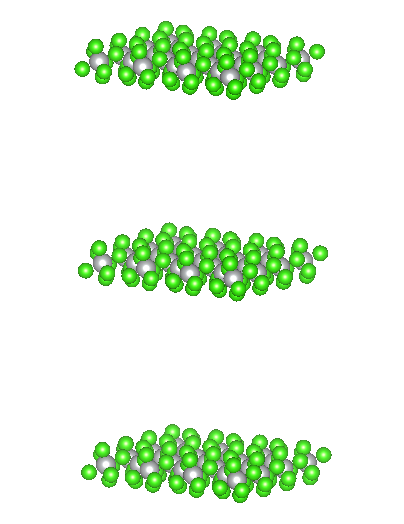}
        \subcaption{NiCl$_4$}
    \end{minipage}
    \begin{minipage}[c]{0.24\textwidth}
        \centering
        \includegraphics[width=\textwidth]{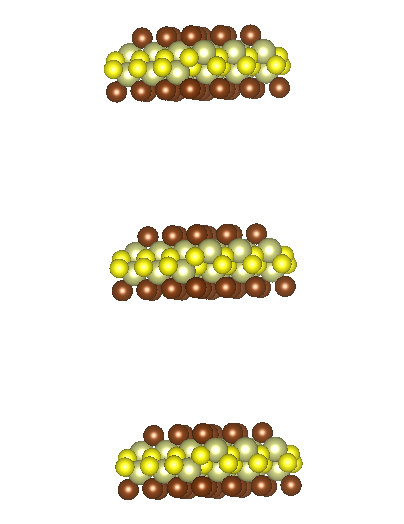}
        \subcaption{IrSBr}
    \end{minipage}
    \begin{minipage}[c]{0.24\textwidth}
        \centering
        \includegraphics[width=\textwidth]{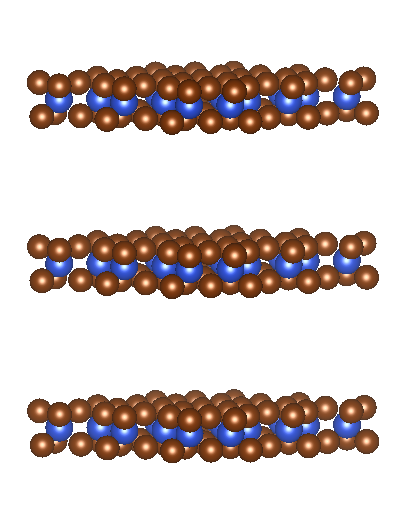}
        \subcaption{CuBr$_3$}
    \end{minipage}
    \begin{minipage}[c]{0.24\textwidth}
        \centering
        \includegraphics[width=\textwidth]{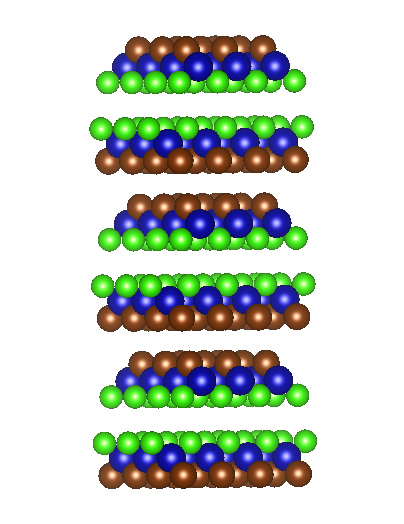}
        \subcaption{CoBrCl}
    \end{minipage}\\
    \caption{Four new 2D structures discovered by our MTG pipeline with 0 E-above-hull energy.}
    \label{fig:E-above-hull}
\end{figure}

Figure\ref{fig:E-above-hull} shows four new 2D structures discovered through our MTG pipeline that have 0 e-above-hull energy (See cif information in Supplementary file). All the structures show a layered structure with each layer forming a compact 2D structure, demonstrating that they have passed the DFT stability check and the capability of our generative 2D materials design pipeline to find new 2D materials.

\FloatBarrier

\section{Conclusion}

Two-dimensional materials have wide applications due to their unique properties. Here we propose a generative design pipeline for 2D materials discovery by integrating a transformer-based 2D material composition generator, two template-based crystal structure predictors, and a graph neural network potential-based structure relaxation algorithm. It is found that the transformer composition generator can capture the composition preference which allows it to generate chemically valid potential 2D materials. We have applied our 2D generator pipeline to discover four hypothetical 2D materials with e-above-hull energy less than 0. Our pipeline is generic and can be used to train other types of materials' generative design models. 

\section{Code Availability}

The BLMM material composition generator is accessible at \href{https://github.com/usccolumbia/blmm}{https://github.com/usccolumbia/blmm}. The TCSP template-based structure predictor can be accessed at \href{http://materialsatlas.org/crystalstructure}{http://materialsatlas.org/crystalstructure}. The CSPML template-based structure predictor can be found at \href{https://github.com/Minoru938/CSPML}{https://github.com/Minoru938/CSPML}. The BOWSR structure relaxation algorithm can be found at \href{https://github.com/materialsvirtuallab/maml}{https://github.com/materialsvirtuallab/maml}. M3GNet relaxation module can be downloaded from \href{https://github.com/materialsvirtuallab/m3gnet}{https://github.com/materialsvirtuallab/m3gnet}. 

\section{Data Availability}

The V2DB database is available at Harvard Dataverse \href{https://doi.org/10.7910/DVN/SNCZF4}{https://doi.org/10.7910/DVN/SNCZF4}. The C2DB database is available at \href{https://cmrdb.fysik.dtu.dk/c2db/}{https://cmrdb.fysik.dtu.dk/c2db/}. The MC2D database is available at \href{https://www.materialscloud.org/discover/mc2d/dashboard/ptable}{https://www.materialscloud.org/discover/mc2d/dashboard/ptable}. The 2DMatPedia database is available at \href{http://www.2dmatpedia.org/}{http://www.2dmatpedia.org/}.

\section{Contribution}
Conceptualization, J.H.; methodology,R.D., J.H. Y.S.,E.S.; software, R.D., Y.S. ; resources, J.H.; writing--original draft preparation, R.D., E.S., J.H.; writing--review and editing,  J.H; visualization, R.D. and E.S.; supervision, J.H.;  funding acquisition, J.H.

\section*{Acknowledgement}
The research reported in this work was supported in part by National Science Foundation under the grant and 1940099 and 1905775. The views, perspectives, and content do not necessarily represent the official views of the NSF.

\bibliographystyle{unsrt}  
\bibliography{references}

\end{document}